\documentclass[10pt]{article}
\usepackage{amsmath, amsfonts, amsthm, amssymb}
\usepackage{anysize,graphicx,natbib,fancyhdr}
\usepackage{algorithm} 
\usepackage{algpseudocode}
\usepackage{mathtools}
\usepackage{caption}
\usepackage{authblk}
\newcommand{\supp}{Supplementary Information}

\newcommand{\figdir}{.}

\usepackage{natbib}
\usepackage[margin=1in]{geometry}
\title{\bf Global labor flow network reveals the hierarchical organization and dynamics
  of geo-industrial clusters in the world economy}

\author[a]{Jaehyuk Park\footnote{Jaehyuk Park and Ian Wood contributed equally to this work.}}
\author[*a,b]{Ian Wood}
\author[a]{Elise Jing}
\author[a,c]{Azadeh Nematzadeh}
\author[b]{Souvik Ghosh}
\author[b,d]{Michael Conover\footnote{Corresponding Author: conover1618@gmail.com}}
\author[a]{Yong-Yeol Ahn\footnote{Corresponding Author: yyahn@iu.edu}}

\affil[a]{\small Indiana University, Bloomington, IN 47408, USA} 
\affil[b]{\small LinkedIn, Sunnyvale, CA 94043, USA} 
\affil[c]{\small S\&P Global, New York, NY 10004, USA} 
\affil[d]{\small Workday, Inc., Pleasanton, CA 94588, USA\vspace{-5ex}}

\date{\tiny}

\begin{document}

\maketitle

\begin{abstract}
  Groups of firms often achieve a competitive advantage through the formation of
  geo-industrial \emph{clusters}.
  Although many exemplary clusters, such as Hollywood or Silicon Valley, 
  have been frequently studied, systematic approaches to identify and 
  analyze the hierarchical structure of the geo-industrial clusters 
  at the global scale are rare.
  In this work, we use LinkedIn's employment histories of more than 500 million 
  users over 25 years to construct a labor flow network of over 4 million firms 
  across the world and apply a recursive network community detection algorithm 
  to reveal the hierarchical structure of geo-industrial clusters. We show that 
  the resulting geo-industrial clusters exhibit a stronger association between 
  the influx of educated-workers and financial performance, compared to existing 
  aggregation units. Furthermore, our additional analysis of the skill sets of 
  educated-workers supplements the relationship between the labor flow of 
  educated-workers and productivity growth. We argue that geo-industrial clusters 
  defined by labor flow provide better insights into the growth and the decline 
  of the economy than other common economic units.
\end{abstract}

Why are the leading internet companies located near each other
in Silicon Valley? Why do aspiring actors who dream of stardom move to Hollywood?
Even though modern telecommunication technologies allow remote collaboration and 
many companies are no longer restrained by physical supply chains,
numerous and conspicuous geo-industrial clusters concentrate
within small geographical areas. Such geographical agglomeration of interconnected 
firms, or ``clusters''~\citep{porter_location_1998, porter_location_2000},
is a key conceptual framework for policymakers and
business economists, from global organizations such as the OECD~\citep{temouri_cluster_2012},
and the World Bank~\citep{yusuf_growing_2008, zhang_competitiveness_2010} 
to regional development agencies in national governments~\citep{martin_deconstructing_2003}.

However, existing studies on geo-industrial clusters struggle with
the following limitations.
First, the concept of the geo-industrial cluster
is vague, and the considered range of spatial and industrial proximity greatly 
varies across studies~\citep{martin_deconstructing_2003}. 
The lack of a concrete definition hampers the systematic analysis of empirical data,
as well as the creation of a solid policy model. Moreover, the lack of consensus in 
the definition
and the lack of extensive empirical data limits most studies to a small number of cases
~\citep{whittington_networks_2009, eisingerich_how_2010, eriksson_localized_2011,
huber_clusters_2012} and encourages reliance on a top-down approach, 
in which scholars or policymakers subjectively, although with expertise, 
assign an industrial sector code to a set of selected administrative districts
~\citep{spencer_clusters_2010}.
Since clusters arise from the strategic decisions
of firms~\citep{porter_location_1998, porter_location_2000}, 
this top-down approach based on predefined industrial and regional 
codes may fail to capture the organic and emergent nature of clusters and their dynamics.
Furthermore, the connections between clusters have been overlooked as well
~\citep{porter_location_1998, porter_location_2000, whittington_networks_2009, eriksson_localized_2011}.
Here, we reveal the hierarchical organization of geo-industrial clusters across multiple scales
in the global economy and argue that examining their inter-connected, hierarchical structure is 
a critical step towards understanding their role in broader economic contexts.

Our approach to identify geo-industrial clusters and
their hierarchical organization involves identifying concentrated labor flow between firms  (see Fig.~\ref{fig:lfn}A). 
The job transitions of workers, labor flow, is central in driving firms
to form geo-industrial clusters due to 
knowledge spillover and labor market pooling~\citep{delgado_clusters_2010,
stephen_tallman_knowledge_2004, agrawal_gone_2006}.
Labor flow thus provides crucial clues to the identification of geo-industrial clusters.
To map these geo-industrial clusters we leverage LinkedIn's dataset which
documents the professional demographics 
and employment histories of more than 500 million individuals 
between 1990 and 2015, allowing us to create, to our knowledge, the largest global \emph{labor flow network}
~\citep{tambe_job_2013, guerrero2013laborflow, neffke2017interindustry} yet 
analyzed. 
The network consists of directed, weighted edges capturing 
approximately 130 million job transitions between more than 4 million firms. 
We show that the structure of this global labor flow network reveals the
multi-scale hierarchical organization of geo-industrial clusters, which constitute a natural,
emergent unit of analysis for the global economy.  

\section{Results}
\begin{figure*}[htpb] 
    \centering
    \includegraphics[width=0.9\linewidth]{\figdir/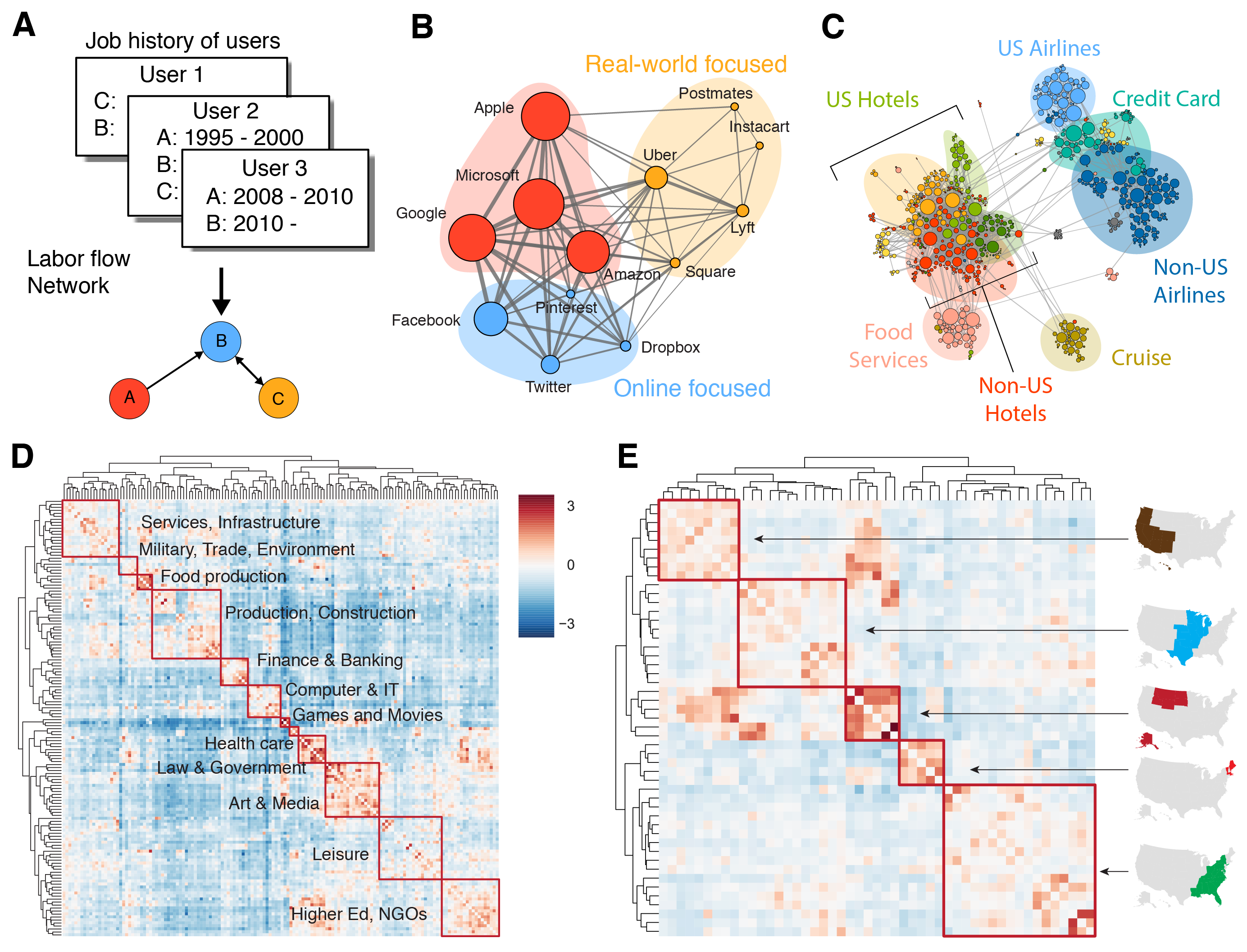}

\caption{The structure of the labor flow network is shaped by firms' geographic 
  proximity and talent demands. 
\textbf{A}, A labor flow network is comprised of organizations (nodes) and 
the flows of people between them (directed, weighted edges) as defined by 
historical records of job changes. 
\textbf{B} and \textbf{C}, Two illustrative examples of geo-industrial clusters defined 
as hierarchically-organized geo-industrial clusters in the labor flow network 
with high intra-cluster talent mobility. 
\textbf{B}, Within a cluster of software / internet technology firms, 
we see sub-clusters with respect to types of services --- online 
(blue), offline (yellow), and both (red), which are also linked to the age of firms.
\textbf{C}, Microindustries shaped by geography and area of specialization are 
also evident within
a cluster of travel-related firms.  
\textbf{D} and \textbf{E}, A transition matrix of labor flows between LinkedIn users' 
self-reported industries (normalized with the expected transition volume) 
highlights labor flows within and between macroeconomic sectors
(see Figure S1 for all original industry labels). 
The effect of geographic proximity on labor mobility is also evident in 
the matrix of labor flows between US states (see Methods for details)
}\label{fig:lfn}
\end{figure*}

Workers tend to change their jobs between geographically 
close firms with similar skill requirements
~\citep{bjelland2011employer, van2001spatial, zipf1946p, simini2012universal}.
This tendency leads to knowledge spillover and innovation, serving as a 
prominent feedback mechanism in the formation of 
geo-industrial clusters~\citep{almeida_localization_1999, cooper_innovation_2001,
moen_is_2005, eriksson_localized_2011, poole_knowledge_2012}.
As geo-industrial clusters form, they further constrain labor flow,
creating a strong concentration of similar skills and experience locally. 
This feedback produces concentrated job migration, which 
in turn can be leveraged to identify 
clusters as \emph{network communities}, groups of 
cohesively interconnected nodes on a network
~\citep{girvan2002community, fortunato2010community}; 
in a labor flow network, the cluster of firms would manifest
as network communities, tied together by concentrated labor flow (see Fig.~\ref{fig:lfn}).  

From our data, relevant geo-industrial clusters can easily be found across domains, from technology firms 
of distinct flavors and ages (Fig.~\ref{fig:lfn}B) to 
clusters of travel and hospitality industries (Fig.~\ref{fig:lfn}C), which 
are concentrated with respect to both specialization (e.g. airlines, promotional 
credit cards, food service, or cruise lines) and geography. 
The hierarchical structure of these geo-industrial clusters is evident in the makeup of 
the non-US airline geo-industrial cluster, which, itself, is comprised of smaller 
sub-modules corresponding to serving geographically distinct markets such 
as Europe and the Middle East. 

The concentration based on the industrial and geographic proximity
can be separately observed through an industry-wise and 
a region-wise transition matrix.
We calculate two normalized transition matrices between industries 
and U.S.~states respectively (Fig.~\ref{fig:lfn}D, E; see Methods for details). 
Industries are split into two large clusters, which roughly correspond to 
production (upper left) and public and consumer services (bottom right). 
In the context of the three-sector theory~\citep{fisher1939production, 
clark1967conditions}, or rather a more recent four-sector framework~\citep{kenessey1987primary}, 
the upper-left cluster is organized around the 
primary, secondary, and some of the tertiary sector (infrastructure and 
business support), while the bottom-right cluster consists of industries mostly 
in the quaternary sector, including higher education, government, law, health 
care, leisure, and media.
Although finance and information technology are often classified into the 
quaternary sector, here they are clustered with production and manufacturing, 
highlighting their strong connection to engineering and production. 
Retail, on the other hand, is clustered more closely with other quaternary 
services, as opposed to tertiary services. 

The abundance of off-diagonal interactions emphasizes the interconnected 
nature of the economy. 
For instance, the law and government sectors are more likely to generate
a cluster with military, trade, and environment sectors than other sectors of the economy, 
although such connections cross the boundary of the two largest industry clusters. 
Curiously, the leisure industry is one of the most widely connected, 
exhibiting strong connections to many other sectors, including healthcare, 
education, art, media, and manufacturing. 
The labor flow network also displays strong geographical clustering, 
as shown in Fig.~\ref{fig:lfn}E. 

\begin{figure*}[htpb] 
 \centering
\includegraphics[width=.7\columnwidth]{\figdir/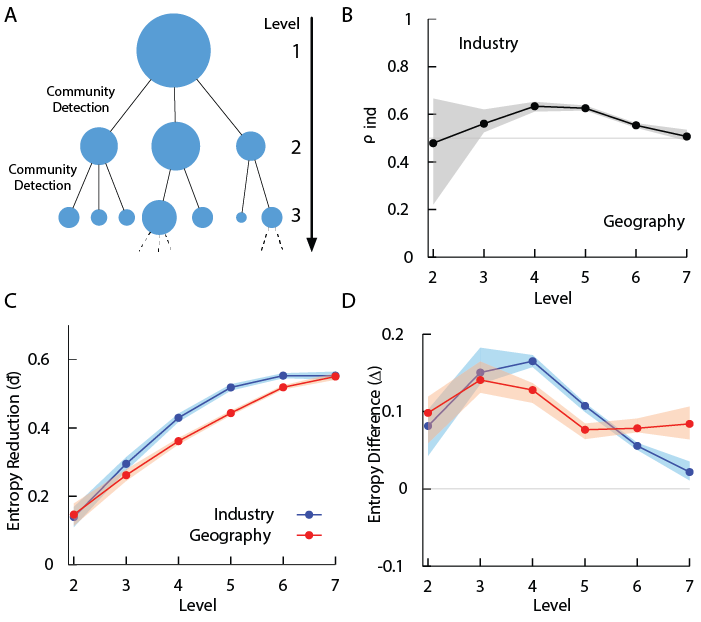}

\caption[Hierarchical structure of the labor flow network]{The impact of
geography and industry across scales. 
\textbf{A}, We recursively apply network community detection to discover the 
labor flow network's hierarchical structure. 
See Methods for more details. 
\textbf{B} Both industry and geography affect job transition across all scales, 
but industry plays a more important role in the middle of the hierarchy as seen 
by the proportion of communities with a greater reduction in industry entropy 
($\rho_\mathrm{ind}$). 
\textbf{C} The average reduction of metadata entropy $\bar{d}$
(see Methods for the definition) at each level of the hierarchical community
structure, calculated with respect to the whole network.
The mononotic increase indicates that smaller communities are more homogeneous as expected. 
\textbf{D} This entropy reduction is greater than expected by a null model. 
$\Delta$ is the difference between the observed entropy reduction and the 
reduction in a randomized hierarchical null model (see Methods and \supp
for more details). Positive $\Delta$ indicates that the homogeneity of clusters 
are stronger than expected.}\label{fig:hierarchy}

\end{figure*} 

The clear presence of clustering with respect to both industry and geography 
prompts the following questions: which factor is more important in 
determining the structure of geo-industrial clusters, industry or geography? 
How do these factors shape the hierarchical structure of these clusters? 
If the composition of a geo-industrial cluster is heavily constrained by industrial or 
geographical proximity, 
we expect to see clusters form around an industry or a location, respectively. 
Therefore, measuring cluster homogeneity in terms of industry and region not 
only allows us to evaluate the validity of clustering but also allows 
us to estimate the strength of each constraint. 
In doing so, we assess the relevance of the clusters as well as the strength of 
industrial or geographical constraints.

We quantify the homogeneity of network communities by calculating the Shannon 
entropy of \emph{cluster feature vectors} that document the fraction of people 
in the geo-industrial cluster who belong to each industry or region (see Methods). 
We quantify the relative importance of industry and geography by calculating 
the ratio between the number of geo-industrial clusters at each level with a greater 
reduction in industrial entropy and those with a greater reduction in geographical entropy.
Our measurement in Fig.~\ref{fig:hierarchy}B-C shows that the industry tends to 
play a more important role than geography in constraining labor flow and its 
strength is strongest at the middle of the hierarchy. 
In other words, network communities tend to be broken down into smaller 
communities mainly based on industrial categories. 
As shown in Fig.~\ref{fig:hierarchy}D, the average entropy reduction is larger 
than expected by chance throughout the hierarchy, indicating that the
identified clusters are cohesive and meaningful.
Then, how are they organized within the global network? 

\begin{figure*}[htpb] 
    \centering
\includegraphics[width=.9\linewidth]{\figdir/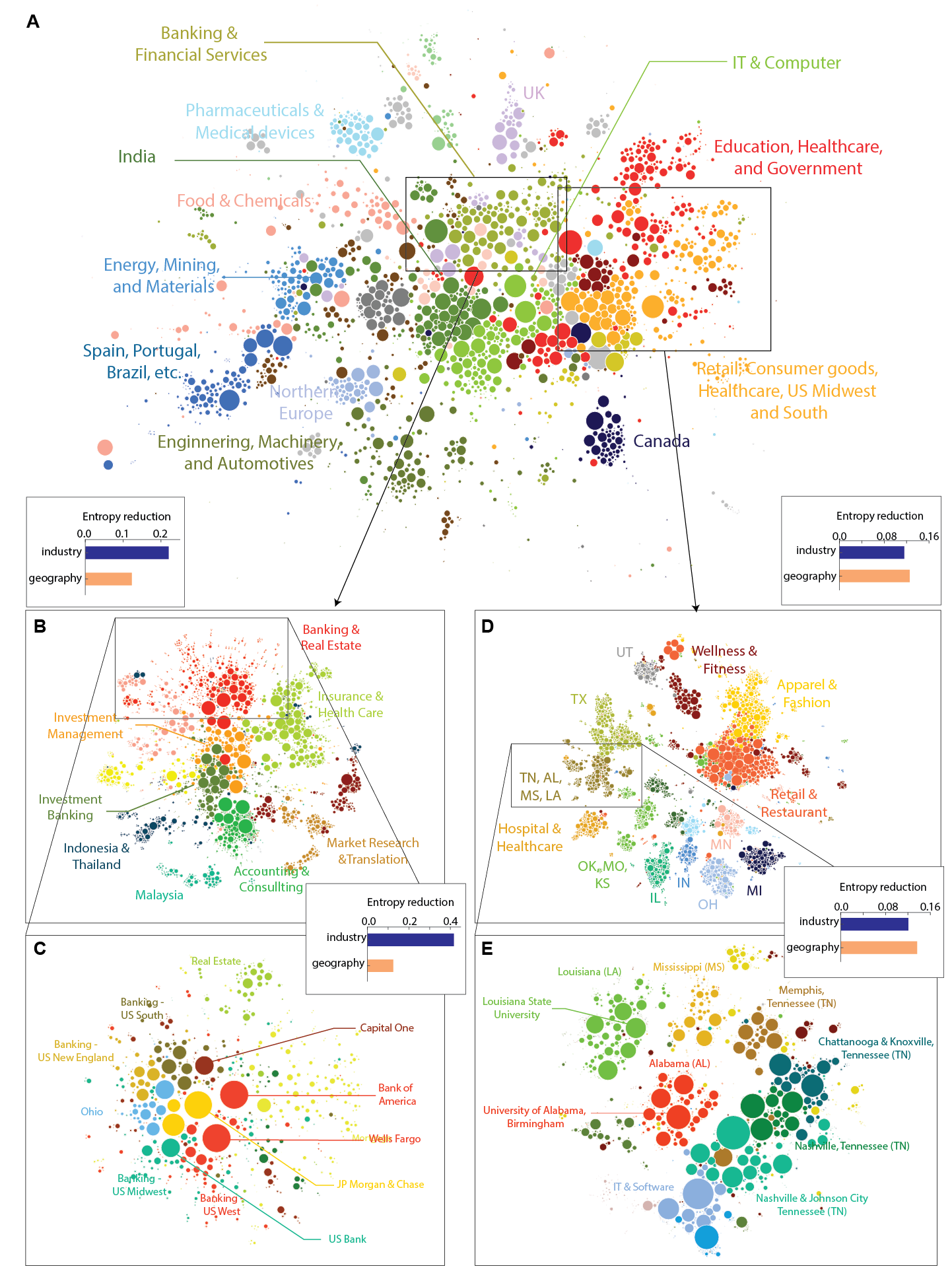}

\caption[Example of hierarchical structure]{\textbf{A}, The large-scale 
organization of geo-industrial clusters in the labor flow network.  
Each circle represents a geo-industrial cluster, with size proportional to its number of employees.  
The colors represent the highest-level community membership. 
\textbf{B--E} Two examples of hierarchical sub-structures in the labor flow 
network are illustrated. 
Each circle represents a firm and the bar charts show the reduction in 
industry and region entropy within the cluster as a proportion of the parent cluster's entropy. 
\textbf{B, C}, the organization of banking and financial geo-industrial clusters are 
affected more by industries than geography. 
\textbf{D, E}, the geo-industrial clusters in U.S. Midwest and South region form strong 
geographical clusters.}\label{fig:hier_network_viz}
\end{figure*} 

We visualize the network of geo-industrial clusters in Fig.~\ref{fig:hier_network_viz}A 
(see Methods for details), where each circle represents a geo-industrial cluster, colored 
based on the highest-level community membership. 
We label each highest-level cluster based on the dominant industry or 
geographical region (See Methods).  
The map exhibits both industry- and geography-dominated clusters.  
Cultural and regional economic blocs, such as Northern Europe, are visible, 
while industrial clustering is also evident. 
For instance, engineering and machinery are associated with automotive 
clusters, and food production and chemicals are associated with 
pharmaceutical and medical devices.  
The map also reveals geographical specializations. 
Firms located in the Midwest of the United States closely interact with retail 
and consumer goods industries worldwide, while India-based clusters are 
strongly associated with information technology.

Zooming into lower levels of the geo-industrial hierarchy reveals more intricate 
structures (See Fig.~\ref{fig:hier_network_viz}B-E). 
Two high-level clusters are shown: one focused on banking and financial 
services in the U.S., and the other with higher education, health care, 
and retail industries in the U.S. 
The banking and financial cluster is broken into more specific industries, 
such as investment banking and real estate (Fig.~\ref{fig:hier_network_viz}B). 
The entropy reduction measure confirms that this hierarchical structure is 
dominated by industrial categories rather than geographical clustering. 
On the other hand, the Higher Education, Health Care, and Retail cluster is 
mostly divided along regional lines. 
These examples depict the structure of the labor flow network as a complex 
tapestry of industry and geography. 

If geo-industrial clusters can effectively capture both 
industrial and geographical proximity, can they serve as a useful framework 
to study the effects of strategic advantage on economic performance?  
The competition for highly desirable jobs implies that well-educated individuals 
who are equipped with strong skill sets would be attracted to the sectors and 
regions that can pay premium wages or rapidly growing ones that may in the future.
Furthermore, the industries and regions that attract well-educated people are 
more likely to benefit from accumulated human capital and spillover 
effects~\citep{pennings1998human, hitt2001direct, simon2002human, chen2005empirical, 
shapiro2006smart, florida2008inside, moretti2010local, moretti2012new}. 
Motivated by these studies as well as a study on the effect of labor market
integration and knowledge spillover within geo-industrial clusters
~\citep{delgado_clusters_2010, stephen_tallman_knowledge_2004, agrawal_gone_2006},
we examine the labor flow of college-degree workers across regions, industries, 
and geo-industrial clusters.

\begin{figure*}[htpb] 
\centering
\includegraphics[width=0.9\linewidth]{\figdir/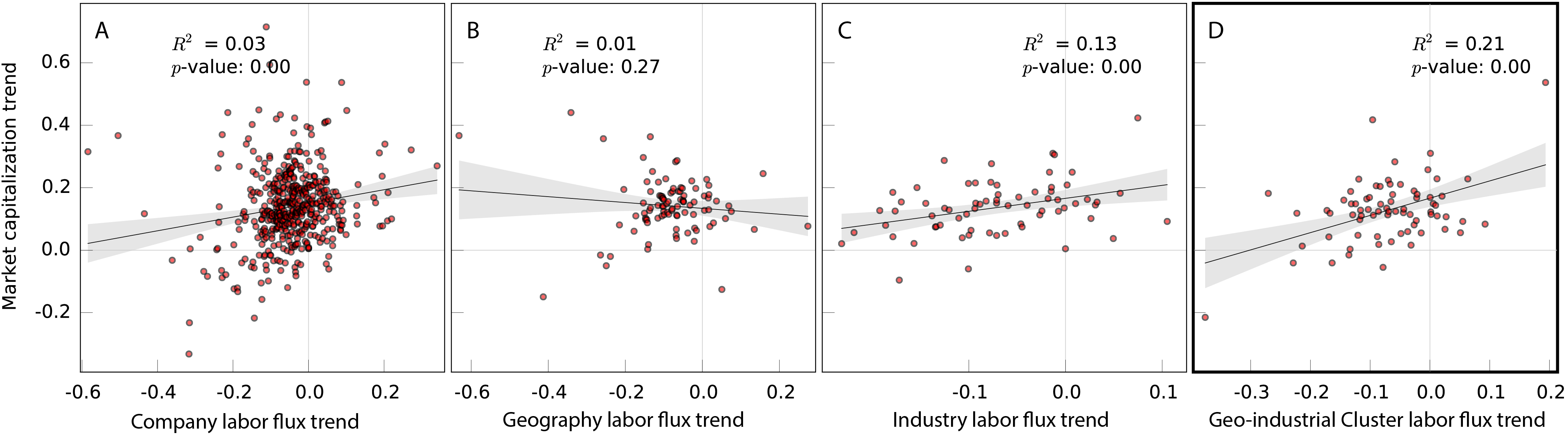}
\small
\caption{The influx of educated labor force is linked to the growth of geo-industrial clusters. 
The horizontal axis represents the five-year trend in college-degree
labor flux from 2010 to 2014. 
Similarly, the vertical axis represents the five-year trend in log-scaled 
market capitalization within the cluster over time. 
\textbf{A} The trends for individual firms, \textbf{B} The trends for 
geographical regions, \textbf{C} The trends for industries, and \textbf{D}.
The trends for geo-industrial clusters, which displays the strongest relationship.}
\label{fig:trends}
\end{figure*}   

We test how well the influx of educated labor correlates with financial performance 
when aggregated into different units of analysis. focusing on the firms in the 
S\&P 500 Index and a time window between 2011 and 2014, 
we compare their market capitalization growth —-- measured by the linear temporal 
trend of log-scaled market capitalization —-- to the labor flux growth —-- 
measured by the linear temporal trend of the log ratio of college-degree labor 
influx to outflux aggregated in each grouping (see Figure~\ref{fig:trends} and Methods).

Overall, we see a positive relationship between the acceleration of college-degree 
employment growth and market capitalization growth although the strength of the 
relationship differs depending on the aggregation used (see Figure~\ref{fig:trends}). 
At the level of individual firms, the data is too noisy to establish any clear 
patterns (Figure~\ref{fig:trends}A). Geographical aggregation similarly shows 
little association between labor growth and market capitalization growth 
suggesting that location-based grouping is also not a good approach, 
probably because each location hosts a multitude of disparate industries. 
Although the industry-level aggregation in Figure~\ref{fig:trends}C shows 
a stronger relationship, the strongest correlation can be found in the 
geo-industrial cluster-based aggregation (see Figure~\ref{fig:trends}D). 
These results hold for more complex bayesian models and are robust to the 
selection of time window, or the inclusion or exclusion of first-job influx 
and last-job outflux (see Supplementary Information). 
The stronger association between the influx of educated labor and economic 
growth in the geo-industrial cluster level, in comparison with traditional 
industry- or region-based aggregation, suggests that firms that share labor 
also share economic growth or decline. This is perhaps due to shared competitive 
advantages due to labor market integration and knowledge spillover effects~
\citep{porter_location_1998, porter_location_2000, delgado_clusters_2010, 
stephen_tallman_knowledge_2004, agrawal_gone_2006}.

\begin{figure*}[htpb] 
    \centering
    \includegraphics[width=\linewidth]{\figdir/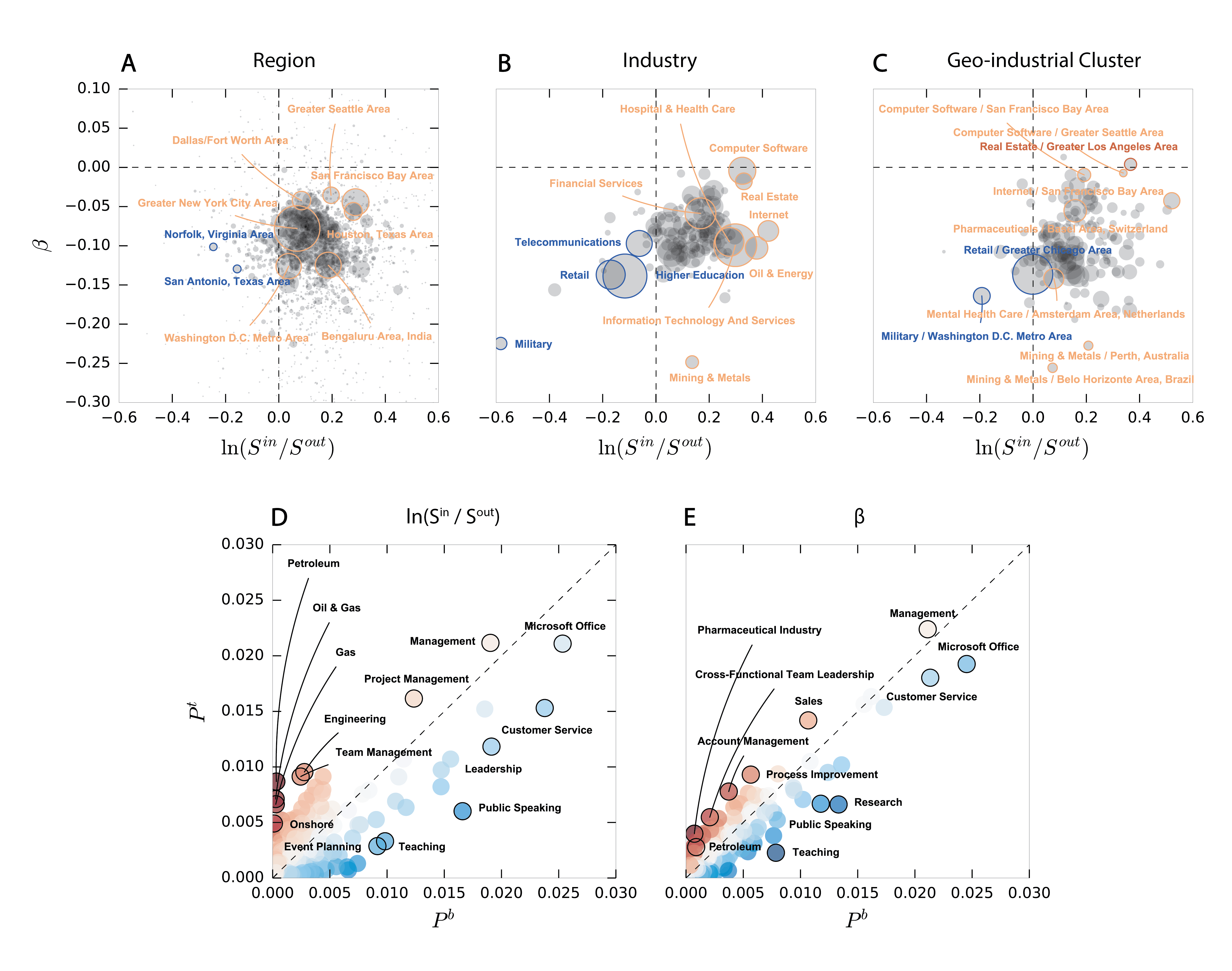}
\caption[Growth of Regions, Industries, and Geo-industrial Clusters and associated 
skills in Growing and Declining Geo-Industrial Clusters]{\textbf{A-C} The log-ratio of 
influx to outflux and its growth over time, aggregated by region (\textbf{A}), 
industry (\textbf{B}), and geo-industrial cluster (\textbf{C}). 
The amount of growth is calculated by the log-ratio of influx to outflux 
($\log(S^\mathrm{in}/S^\mathrm{out})$) during each year from 2010 to 2014; its linear time 
trend ($\beta_i$) is estimated by the linear regression coefficient of influx 
ratios to time over this period. 
The size of a circle represents the number of total transitions either into or 
out of a corresponding category.  
\textbf{D, E} Overrepresented skills in geo-industrial clusters in the top and bottom 
quartiles of log-ratio influx to outflux (\textbf{D}) and its linear time 
trend (\textbf{E}).  
The fraction of people who have a certain skill in the top ($P_{q}^t$) and 
bottom ($P_{q}^b$) geo-industrial clusters reveals that specialized and business-oriented 
skills are more common in growing geo-industrial clusters than declining 
geo-industrial clusters.}\label{fig:skill}
\end{figure*}

We see that the influx of educated workers to a geo-industrial cluster is a 
meaningful signal of growth, so we can ask which regions, industries, and 
geo-industrial clusters are seeing that growth. we measure the total growth 
in terms of influx during a period from 2010 to 2014, using the log ratio of 
influx to outflux of college-educated workers for each region, industry and 
geo-industrial cluster, $\log(S^\mathrm{in}/S^\mathrm{out})$ 
(See Figure~\ref{fig:skill}A-C and Methods). 
We then estimate the change of this growth, denoted $\beta$, 
by estimating the linear trend in time of the influx log-ratio during 
the same period. If a region, an industry, or a geo-industrial cluster exhibits 
a positive net influx and a positive $\beta$, 
it means that it has been growing and the growth has been increasing 
during this period.

Figure~\ref{fig:skill}A shows that most regions are located in the fourth quadrant,
with decelerating growth following a strong bounce-back from the Great Recession 
of 2007-09~\citep{cunningham_great_2018}.
The San Francisco Bay area and the Greater Seattle Area exhibited the
strongest growth, while places such as San Antonio have been 
losing educated population. 
Similarly, most industries also show a slowing growth out of the recession 
(see Fig.~\ref{fig:skill}B). In this period, the ``Computer Software'' industry has been 
showing the strongest growth, while ``Retail'' has been losing its educated 
labor force. This trend has been accelerating. 
Also note that the ``Mining \& Metals'' industry has been growing but decelerating, 
and the ``Internet'' and ``Oil \& Energy'' industries experienced large growth 
during this period. 
These employment growth patterns match the relative growth projections from 
the U.S. Bureau of Labor Statistics' Occupational Handbook~\citep{occupationalhandbook}, 
except that our analysis detects a loss of ``Retail'' jobs among the college-educated,
and a pronounced deceleration in growth across many fields.  

Although these region- and industry-based views paint a rough picture that fits 
the known recent trends of the global economy, it is the geo-industrial cluster-based 
analysis that provides the best snapshot of the evolution of the economy. 
The fact that the San Francisco Bay area has been rapidly growing does not tell 
us \emph{which industry} propelled the growth; likewise, the growth of the computer 
software and internet industries does not inform us \emph{where} this growth has occurred.
By contrast, a cluster-based comparison in Fig.~\ref{fig:skill}C reveals
nuanced information about the growth of geo-industrial clusters, completing 
the picture of economic evolution during this period.  
The clusters that are based on internet and computer software companies 
in the San Francisco area, real estate companies in the Los Angeles area, and 
computer software companies in the Seattle area experienced some of the strongest 
growth with respect to college-degree workers, while military-related firms and 
organizations in Washington D.C. and retail companies in the Chicago area experienced 
the largest decline. 

This pattern of productivity growth can be supplemented with an even more 
detailed analysis of associated skills. 
Here, we identify over- and under-represented skills in emerging and declining 
geo-industrial clusters.
We compare the aggregated skill distribution of geo-industrial clusters in the top-quartile 
of total influx ($\log(S^\mathrm{in} / S^\mathrm{out})$; Fig.~\ref{fig:skill}D) or growth 
($\beta$; Fig.~\ref{fig:skill}E) during this period against those in the bottom quartile.  
The vertical axis represents the fraction of employees with each skill within 
the top quartile, and the horizontal axis represents the proportion in the bottom quartile. 
The intensity of the color represents the degree to which each skill is concentrated 
in the top (red) or bottom (blue) quartile, as measured by the z-score of the 
log-odds ratio between the top and bottom skill distributions (see Methods).  
With respect to the total influx, the over-represented skills in the top 
geo-industrial clusters are concentrated around management skills, such as ``management'',
``project management'', and ``team management''. 
These results concur with studies on the importance of cognitive-social 
skills and the prevalence of management-related jobs in high-wage occupations
~\citep{autor_putting_2013, autor_skills_2014, deming_growing_2017}. 
In addition, oil and energy-related skills such as ``petroleum'',
``oil \& gas'', ``gas'', and ``onshore'' are more prevalent in the top quartile, 
which captures the recent growth of oil and natural gas industry, driven by the 
new drilling and fracking technologies applied in the U.S. during this 
period~\citep{rampton2012reuters, zakaria2012times, brown2013cfr, plumer2014vox}. 

On the other hand, the most over-represented skills in geo-industrial clusters in the bottom quartile 
feature widely-available, common skills such as ``customer service'' 
and ``Microsoft Office'', or vague skills such as ``leadership''. 
This bias towards common and vague skills in the bottom quartile remains 
consistent regardless of the focus on the total influx or its growth (Fig.~\ref{fig:skill}E). 
Although the ``leadership'' skill is more common in the bottom quartile, 
related, but more specific skills, such as ``cross-functional team leadership'' 
or ``process improvement'' are over-represented in the top growing geo-industrial clusters. 
The over-represented skills in the top quartile of influx \emph{growth} feature 
newer skills, such as ``pharmaceuticals'', ``biotechnology'', and ``cloud computing'', 
capturing new innovations that are attracting educated labor flow. 

\section{Discussion}
In this study, we propose a systematic approach to identify geo-industrial clusters
by analyzing a massive dataset from LinkedIn that captures 
individual-level labor flow between firms across the world. 
The map of our geo-industrial clusters is generated organically  
by high-resolution individual-level data, and allows us 
(1) to identify the geo-industrial clusters systematically through
 network community detection,
(2) to verify the importance of region and industry in labor mobility,
(3) to compare the relative importance
between the two constraints in different hierarchical levels, and
(4) to reveal the practical advantage of the geo-industrial cluster
as an unit of future economic analyses.

At the same time, we would also like to note a number of caveats and limitations of our study.  
Although LinkedIn is widely adopted across the world, the population is still 
biased towards the U.S. as well towards a younger population with more technical backgrounds. 
Moreover, the adoption of LinkedIn is likely to be affected by social diffusion
processes, so its data may exhibit stronger clustering and uneven biases. 
In addition, our approximation uses each firm as a homogeneous unit, which may 
be inadequate, particularly for large firms that host a wide variety of jobs
that are not directly connected to the firms' main products. 
Also, we assume geo-industrial clusters are disjoint sets although they are likely to  overlap in real world.
Finally,
our results on the correlation between labor concentration and 
market capitalization growth are
not enough to prove that the influx of educated workers \emph{leads} to higher 
valuation because there may exist other confounding factors, or the direction 
of causality may be the opposite --- higher valuation leads to more hiring
of educated workers. 
Additionally, this analysis focuses only on S\&P 500 firms and thus should be 
interpreted carefully. 

We argue that, even with these caveats, the labor flow network approach 
can provide powerful and novel ways to examine how economies are organized and evolve. 
Because we focus on the flow between firms, industries, and regions, rather than 
their size, our results show enough consistency to overcome representation biases.
For instance, we expect that the transition matrix in Fig.~\ref{fig:lfn} would 
be robust against representation biases unless job transition patterns and 
LinkedIn membership are strongly confounded, and as long as representation 
bias does not strongly alter the differences between intra- and inter-cluster flow. 
Finally, as in a previous study on cultural history~\citep{schich2014history}, 
focusing on an important sub-population may provide more meaningful results.  
Given the high resolution, coverage, and flexibility, we argue that the global 
labor flow network and geo-industrial cluster framework can serve as a basis 
for future economic analysis. 

Our study may provide a foundation for further systematic analysis of geo-industrial
clusters in the context of 
business strategy, urban economics, regional economics, and international development
fields as well as providing useful insights 
for policymakers and business leaders. 
For instance, our methodology can be applied to other similar, smaller scale
datasets to discern clusters within a single category and to examine
their interconnectedness.

\section{Methods}
\subsection{Labor flow network} 
A \emph{labor flow network} is a directed, weighted graph, $G(V,E,W)$, in which 
each node $u \in V$ corresponds to a firm and each edge $e_{i \rightarrow j} 
\in E$ represents the number of individuals who reported employment at firm 
$i$ prior to moving to firm $j$ in a given time period $(t_s, t_e)$. 
A job transition is included if the start of a job at new firm $j$ begins 
after the start of the time period $t_s$ and before the end of the time period 
$t_e$, even if the job at $i$ was begun before $t_s$. 
The weight of each edge $w_{i \rightarrow j} \in W$ corresponds to the total 
number of recorded job transitions from firm $i$ to firm $j$ in the time window. 
If a member reports multiple job transitions ending or beginning in the same 
month (the smallest resolution of our time data) a unit weight is divided into 
all associated transition edges so that $\frac{1}{k}$ is added to each edge, 
where $k$ is the number of edges. 
The size of firm $i$ at time $t$, $s_i(t)$, is defined by the number of 
members who reported working at firm $i$ at $t$.

We constructed a labor flow network utilizing job history data spanning 1990 to 
2015, $G_{(1990, 2016)}$. 
We then apply the following procedures to obtain the core of the network: 
(1) removing edges with $w_{i \rightarrow j} < 2$; (2) 2-core filtering 
(removing dangling nodes); and (3) isolating the largest connected component.  
This process produces a network representing approximately 42 million job 
transitions over 8,319,091 edges between 487,782 firms. 
For yearly analysis, given a year $t$ we create a labor flow networks 
$G_{(t, t+1)}$ performing no further filtering.

The detailed hierarchical structure is identified by recursively applying 
the Louvain community detection algorithm~\citep{blondel2008fast, leicht2008community}.  
We start with the maximum modularity partition and keep applying the same method 
to each community subgraph if the community has more than 10 nodes. 
The hierarchical tree that connects each community to its subcommunities is 
then pruned using metadata, as explained in the following sections.

\subsection{Company and cluster feature vectors} 
Each firm $c$ is characterized by a set of \emph{firm feature vectors}, 
namely a geography vector $\vec f^{(G)}(c)$ and an industry vector $\vec f^{(I)}(c)$.  
Each element of the vectors represents the fraction of employees of firm $c$ 
who reported a particular attribute (i.e.~a specific region or industry) in their profile. 
We define the region (industry) of a firm as the most frequent region (industry) 
in $\vec f^{(G)}$ ($\vec f^{(I)}$).  
Similarly, for a given community of firms, $C$, we can describe a 
\emph{cluster feature vector} $\vec F(C)$ where each element represents the 
fraction of all employees of the firms in the cluster that report that 
particular attribute.

\subsection{Mapping transitions between industry and geographical regions} 
\label{par:industry_flux}
We construct two transition matrices, one representing labor flows between 
industries and another representing transitions between the states in the U.S. 
In these matrices, each element $T_{ij}$ represents normalized transition weight 
from $i$ to $j$ ($i$ and $j$ can be either two industries or two regions). 
The expected flux between $i$ and $j$ is estimated by 
\begin{equation} 
\mathop{\mathbb{E}}(w_{i\rightarrow j}) = S^{\mathrm{out}}_i \frac{S^{\mathrm{in}}_j}{\sum_k S^{\mathrm{in}}_k},
\end{equation}
where $S^{\mathrm{out}}$ is the total number of members who 
moved out of $i$, and $ S^{\mathrm{in}}_j$ is the total number of members that
moved into $j$. 
Thus the normalized flux from $i$ to $j$ is estimated by
\begin{equation} T_{i\rightarrow j} = \frac{w_{i\rightarrow
j}}{\mathop{\mathbb{E}}(w_{i\rightarrow j})}. \end{equation}
As a result, we have $T_{i\rightarrow j}>1$ if there are more people moving
from $i$ to $j$ than expected by the given null model, and  $T_{i\rightarrow
j}<1$ vice versa. 

\subsection{Measuring cluster homogeneity} 
We measure the homogeneity of a cluster using Shannon entropy, a measure of 
specificity defined for industry vectors by $H^{I} (C) = - \sum_{i} F^{I}_i(C) 
\log F^{I}_i(C)$, where $F^{I}_i(C)$ represents the cluster feature vector of 
the geo-industrial cluster $i$, in terms of industry. 
With geographic entropy, $H^G$ defined similarly using $F^{G}_i(C)$, the cluster 
feature vector of the geo-industrial cluster $i$ in geography. 

\subsection{Detecting over-represented labels} 
To identify over-represented industries or geographical regions in a cluster, 
we employ the \emph{log-odds ratio informative Dirichlet prior} 
method~\citep{monroe2008fightin}. 
The log odds ratio of industry or region $w$ in cluster $i$, compared with 
cluster $j$ is 

\begin{equation} 
  \delta^{i-j}_w = \log \left(\frac{f^i_w + f^{b}_w}
  {N^i + N^{b} - (f^i_w + f^{b}_w)} \right) - \log \left( \frac{f^j_w + f^b_w}{N^j 
  + N^b - (f^j_w + f^b_w)} \right) 
\end{equation} 
\normalsize
where $f^i_w$ is the frequency 
of $w$ in cluster $i$, $f^b_w$ is the pseudo-count for $w$ in the Dirichlet 
prior, $N^i$ is the number of labels in cluster $i$, and $N^b$ is the sum of 
Dirichlet pseudo-counts. 
Then the variance and Z-score are estimated as following: 

\begin{equation} \sigma^2 
  (\delta^{i-j}_w) \approx \frac{1}{f^i_w + f^b_w} + \frac{1}{f^j_w + f^b_w}, 
  Z = \frac{\delta^{i-j}_w}{\sqrt{\sigma^2 \left( \delta^{i-j}_w \right)}} 
\end{equation} 
\normalsize
We make an approximation by considering all other clusters as the 
`other' cluster ($j$) and the set of all firms as the background corpus.

\subsection{Metadata-based pruning} 
We employ a metadata-based stopping heuristic for recursive community detection 
to identify a particular partition from the hierarchical structure.  
Our main idea is that (1) we can safely split a community if it can be broken 
into multiple communities, each of which exhibits strongly over-represented 
industry or geographical region metadata, and (2) that such splitting is 
inappropriate if the resulting children do not have any over-represented metadata. 
Our method moves down the tree from the root to the finest level of the community 
hierarchy that maintains significant over-representation of particular regions 
or industries within the community. 
Given two thresholds, $\theta_b$ (break threshold) and $\theta_k$ (keep threshold), 
we look at whether the current community over-represents some industry and 
region label, with Z-score surpassing $\theta_b$. 
If it does not or is a leaf-node, we keep the community if it over-represents 
some industry and region label with a more lenient keep threshold $\theta_k$, 
and otherwise prune the community from the tree. 
If it does over-represent metadata at or above the threshold of $\theta_b$, 
the process is repeated for the community's children. This algorithm is laid 
out in Algorithm \ref{ToMin}.  
We use $\theta_k=1.96$ and $\theta_b=100$ for financial data analysis, as this 
threshold provided a moderate number of communities, without pruning any 
firms for which we had financial data. 
We use $\theta_b=10$ for visualizations. 

\begin{algorithm}
\caption{Pruning To Minimum Threshold}
\begin{algorithmic}[1]

\Require{$tree$, $root$, $\theta_k$, $\theta_b$}
\Ensure{$save\_list$}
	\State $visit\_list \leftarrow root$
	\State $save\_list \leftarrow list()$
	
	\For {$node \in visit\_list$}
		\State $children \leftarrow tree.children(node)$

		\For {$child \in children$}
			\If {$child.max\_Z\_score> \theta_k$ for industry and region}
				\State $visit\_list.append(child)$
			\ElsIf{$child.max\_Z\_score > \theta_b$ for industry and region}
				\State $save\_list.append(child)$
			\EndIf
		\EndFor
	\EndFor
\State \Return{$save\_list$}
	
\end{algorithmic}
\label{ToMin}
\end{algorithm}
  \paragraph{Entropy reduction}  
  We measure the entropy of industry and geographical region cluster feature 
  vectors at each level of the cluster hierarchy to validate our community detection 
  strategy as well as to compare the impact of geography and industry in job transitions. 
  Entropy reduction as shown in Fig.~2 is calculated for both industry and regional 
  labels as a ratio of the difference between the global entropy $H(V)$ and a 
  community $j$'s entropy $H(C_j)$ to the global entropy, $d_j =\frac{H(V)-H(C_j)}{H(V)}$. 
  This ratio is used instead of the raw entropy reduction to provide a comparable 
  scale between industry and geographical region metadata, since there are many 
  more possible region labels than industry labels. 
  $\rho_k = |\{j | d^I_j > d^R_j; j \in K_k\}|/|K_k|$ is the proportion of 
  communities $j$ in the set of communities $K_k$ at level $k$ of the hierarchy 
  with a greater reduction in industry entropy $d^I$ than geographical entropy $d^G$. 
  The average entropy reduction over all communities in each hierarchical level 
  weighted by the number of firms is reported as 
  $\bar{d_{k}} = \frac{\sum_{j \in K_k}{w_j \cdot d_j}}{\sum_{j \in K_k}{w_j}}$ 
  where $w_j$ is the number of firms in community $j$ --- and its standard 
  error is estimated by Cochran's method as reported in~\citep{gatz1995standard}.  
  This is equivalent to the mutual information between community and industry or 
  geography partitions at each level of the hierarchy, normalized by the overall 
  industry or geographical entropy. 
  This is an imperfect measure (and there may be no perfect measure for clustering 
  comparisons~\citep{meila2005comparing}), which still favors comparisons between 
  sets with more possible labels~\citep{vinh2010information}, such that we are 
  likely over-estimating the importance of geography, but it does allow for some comparison. 
  We employ a tree-shuffling null model that randomly shuffles all firms 
  throughout the hierarchical community tree such that the tree is still a 
  consistent community hierarchy; for each firm $c$, a firm $\hat{c}$ is 
  randomly selected from the set of all firms, and $c$ is replaced by $\hat{c}$ 
  firm in each community to which it belonged, giving us corresponding null 
  values $\bar{d_{k}}^{\prime}$ with the difference 
  $\Delta_{k} = \bar{d_{k}} - \bar{d_{k}}^{\prime}$

\subsection{Marketcap trends}  
We use the market capitalization data for S\&P 500 firms from 1996 through 2015. 
For each given partition (i.e.~geographical regions, industries, and selected 
geo-industrial clusters), we aggregate all market capitalization within a cluster by summing them. 
The influx and outflux are also aggregated at the cluster level, ignoring 
within-cluster flow, but including first recorded jobs as influx and last 
recorded jobs as outflux.  
To find trends over time, we performed a ordinary least-squares linear regression 
between a variable representing time and the variable of interest as shown below:
\begin{equation} MC_{i,t} = \beta_{MC,i}t + \mu_{MC,i} \end{equation}
\begin{equation} LF_{i,t} = \beta_{LF,i}t + \mu_{LF,i} \end{equation} 
where $MC_{i,t}$ and $LF_{i,t}$ are the quarter-four log market capitalization and
yearly labor flow respectively for cluster $i$ at time $t$, $\beta$ is the
slope of the regression, and $\mu$ is the intercept. The slope of the
regression is then used as the trend $\beta$ in the further regression:
\begin{equation} 
\beta_{MC,i} = \bold{\beta}_\beta\beta_{LF,i} + \mu_{\beta}
\end{equation}

Although this model is intuitive, it treats inferred parameters as observed. 
A more complete Bayesian model that also accounts for errors in parameter 
estimation is included in Bayesian Model for Trends of Trends in Supplementary Information.

\bibliography{main}

\end{document}